\documentclass[journal]{IEEEtran}
\usepackage[T1]{fontenc}
\usepackage{float}
\usepackage{amsmath}
\usepackage{amsfonts}
\usepackage{mathtools}
\usepackage{amsfonts}
\IEEEoverridecommandlockouts
\usepackage{amssymb}
\usepackage{graphicx}
\usepackage{epstopdf}
\usepackage{makeidx}
\usepackage{csquotes}
\usepackage{graphicx}
\usepackage{cite}
\usepackage{xcolor}
\usepackage{subcaption}
\usepackage{graphicx}
\usepackage{balance}
\begin{document}

\title{Integration of Beyond Diagonal RIS and UAVs in 6G NTNs: Enhancing Aerial Connectivity}

\author{Wali Ullah Khan, \textit{Member, IEEE,} Eva Lagunas, \textit{Senior Member, IEEE,} Asad Mahmood, \textit{Student, IEEE,} Muhammad Asif,  Manzoor Ahmed, Symeon Chatzinotas, \textit{Fellow, IEEE} \thanks{W. U. Khan, Lagunas, A. Mahmood, and S. Chatzinotas are with the University of Luxembourg, Luxembourg (waliullah.khan,eva.lagunas,asad.mahmood,symeon.chatzinotas)@uni.lu; Manzoor Ahmed is with the Hubei Engineering University, Xiaogan, China (manzoor.achakzai@gmail.com); M. Asif is with the School of Computer Science and Communication Engineering, Jiangsu University, Zhenjiang, Chins (masif@ujs.edu.cn).

}}%

\markboth{Accepted in IEEE Wireless Communications Magazine, vo. X, no. Y, 2025
}
{Shell \MakeLowercase{\textit{et al.}}: Bare Demo of IEEEtran.cls for IEEE Journals} 

\maketitle

\begin{abstract}
The reconfigurable intelligent surface (RIS) technology shows great potential in sixth-generation (6G) terrestrial and non-terrestrial networks (NTNs) since it can effectively change wireless settings to improve connectivity. Extensive research has been conducted on traditional RIS systems with diagonal phase response matrices. The straightforward RIS architecture, while cost-effective, has restricted capabilities in manipulating the wireless channels. The beyond diagonal reconfigurable intelligent surface (BD-RIS) greatly improves control over the wireless environment by utilizing interconnected phase response elements. This work proposes the integration of unmanned aerial vehicle (UAV) communications and BD-RIS in 6G NTNs, which has the potential to further enhance wireless coverage and spectral efficiency. We begin with the preliminaries of UAV communications and then discuss the fundamentals of BD-RIS technology. Subsequently, we discuss the potential of BD-RIS and UAV communications integration. We then proposed a case study based on UAV-mounted transmissive BD-RIS  communication. Finally, we highlight future research directions and conclude this work.
\end{abstract}


\IEEEpeerreviewmaketitle

%%%%%%%%%%%%%%%%
%\begin{figure*}[!t]
%\centering
%\includegraphics[width=0.60\textwidth]{BCtypes.eps}
%\caption{Classifications of backscatter communication systems (a) Monostatic system (b) Bistatic co-located system, (c) Bistatic dislocated system, and (d) Wireless-powered ambient system. Here Tx refers to RF carrier emitter, Rx is the receiver, and Tag is the backscatter device. }
%\label{blocky}
%\end{figure*}
%%%%%%%%%%%%%
\section{Introduction}%$\mdlgwhtsquare\mdlgblksquare\boxbox$
Unmanned aerial vehicles (UAVs) are expected to have a significant impact on the advancement and implementation of sixth-generation (6G) non-terrestrial networks (NTNs) due to their versatility, portability, and capacity to function in many conditions \cite{10003076}. They are well-suited to operate in a wide range of applications such as dynamic coverage extension, enhanced Internet of Things (IoT) connectivity, high-altitude relay stations, advanced sensing and data collection, aerial base station (BS), backhaul connectivity, and ultra-reliable low-latency communications (URLLC) \cite{9768113}. Nevertheless, existing wireless communication systems that have limited resources are facing difficulties in satisfying the growing need for a high-quality user experience (QoE). This includes requirements such as fast data rates, minimal delay, and widespread connectivity, which are demanded by the above-discussed applications. In addition, the future deployment of millimeter wave (mmWave)/terahertz (THz) bandwidth in terrestrial and NTNs may encounter several propagation obstacles, including significant signal attenuation and misalignment of transmission beams. In light of this context, reconfigurable intelligent surfaces (RIS) emerge as a promising solution to tackle the aforementioned problems \cite{10584518}. More precisely, RIS can enhance coverage and improve the spectrum and energy efficiency of current wireless networks by controlling the propagation of radio signals in a cost-effective and environmentally friendly manner  \cite{10365519}.

Currently, research on UAV networks predominantly emphasizes conventional RIS, specifically diagonal RIS (D-RIS), which employs diagonal phase shift matrices to manipulate incident signals. In D-RIS, each element independently adjusts the wireless environment without coordination among elements \cite{9703337}. While D-RIS offers a simple design that effectively enhances signal strength, its limited inter-element coordination restricts its beamforming capabilities, thereby reducing its overall performance in more complex communication scenarios. The concept of Beyond Diagonal Reconfigurable Intelligent Surfaces (BD-RIS) has recently been introduced, extending the functionality of reflecting surfaces by enabling a full scattering matrix that is not restricted to a diagonal structure \cite{9514409}. This advanced capability allows BD-RIS to perform sophisticated beamforming, providing simultaneous support for multiple users and facilitating complex signal manipulation. This article explores the advantages of integrating BD-RIS with UAVs, offering enhanced control over the wireless environment through element interconnection \cite{10316535}. In addition to achieving a complete scattering matrix, BD-RIS enables full space coverage by incorporating both signal reflection and refraction \cite{10319662}. This innovative concept has the potential to revolutionize the efficiency and adaptability of UAV networks. To the best of our knowledge, this is the first work to explore the integration of BD-RIS and UAV communication in 6G NTNs.
%%%%%%%%%%%%%%%%
\begin{figure*}[!t]
\centering
\includegraphics[width=0.9\textwidth]{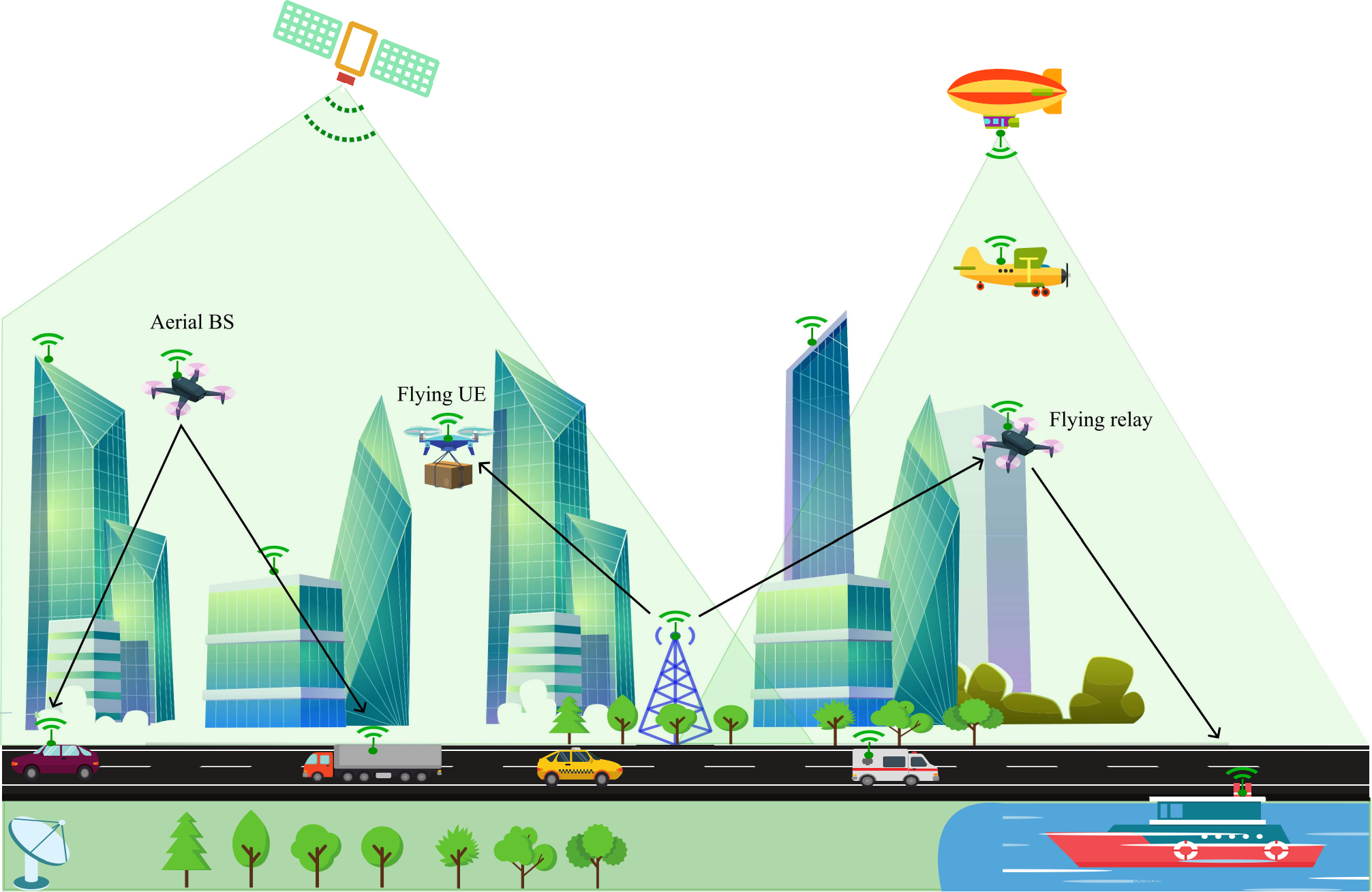}
\caption{Use cases of UAV communications, including flying UE, aerial BS, and relay nodes.}
\label{usecases}
\end{figure*}
%%%%%%%%%%%%%

The main contributions of this article can be summarized as follows. First, we discuss the preliminaries of UAV communications, including applications and use cases of UAV communications in Section \ref{SEC:II}. Then, in  Section \ref{SEC:III}, we present the fundamentals of BD-RIS, discuss the difference between BD-RIS and D-RIS and their key characteristics, analyze the hardware architectures of BD-RIS, and propose various configurations of UAV-based BD-RIS systems. Subsequently, in Section \ref{SEC:IV}, we provide the key potentials when integrating BD-RIS and UAV communications in 6G NTNs. We also provide a technical case study based on transmissive BD-RIS mounted UAV communications and the results are compared with the benchmark transmissive D-RIS mounted UAV communications in Section \ref{SEC:V}. It is shown that BD-RIS outperforms D-RIS in improving the spectral efficiency of multi-user UAV communications. Finally, in Section \ref{SEC:VI}, we highlight future research directions with a concluding discussion in Section \ref{SEC:VII}. 

\section{Preliminaries of UAV Communications} \label{SEC:II}
This section offers an overview of UAV communication applications, followed by an in-depth analysis of use cases where UAVs function as flying user equipment (UE), aerial BS, and relay networks, highlighting their role in strengthening communication infrastructure.
\subsection{Applications of UAVs}
\subsubsection{Enhanced Connectivity}
The upcoming 6G networks are intended to connect massive devices to the internet. UAVs that can function as aerial BSs, relays, or user devices, offer improved connectivity in regions that have limited or nonexistent terrestrial network coverage \cite{9768113}. This is especially beneficial in isolated places, areas affected by disasters, and densely populated urban situations where conventional infrastructure is insufficient.
\subsubsection{Improved System Coverage and Capacity}
Combining terrestrial and non-terrestrial components in 6G networks will augment spectral efficiency and wireless coverage. UAVs can adaptively modify their locations to enhance coverage and capacity. UAVs can establish temporary hotspots to manage increased traffic during crowded events or emergencies, hence improving the network's adaptability and scalability.
\subsubsection{High Reliability and Minimal Delay}
Extremely low latency is a fundamental characteristic of 6G wireless networks. Utilizing UAVs can further reduce delays by establishing more direct communication pathways and bypassing terrestrial limitations. This is crucial for applications requiring real-time responses, such as remote surgery, autonomous driving, and industrial automation.
\subsubsection{3D Network Topology}
Three-dimensional (3D) topology in 6G infrastructure overcomes the constraints of conventional two-dimensional (2D) infrastructure and satisfies the increasing requirements of future wireless networking. In contrast to conventional 2D networks, UAVs provide 3D network topologies, optimizing the utilization of vertical space. This can considerably enhance spectrum efficiency and network performance by optimizing geographical resource allocation.
\subsubsection{Edge Computing and IoT Integration}
Edge computing is critical for 6G systems because it delivers ultra-low latency, improves bandwidth efficiency, enhances privacy and security, and enables real-time processing. By processing data closer to the source and reducing the need for backhaul to central data centers, UAVs can serve as mobile edge computing platforms. This is particularly advantageous for IoT applications, where real-time data processing is essential.
\subsubsection{Integrated Sensing and Communication}
Integrated sensing and communication are integral to the vision of 6G networks, offering significant improvements in user experience, network performance, security, and other applications. Equipped with various sensors and cameras, UAVs can perform various tasks such as data transmission, navigation, and object detection using limited system resources. Integrated sensing and communication capability enables UAVs to simultaneously sense objects in their surroundings while maintaining reliable communication with control centers. This integration improves spectral efficiency, promotes flight safety, and minimizes energy consumption.

\subsection{Use Cases of UAV Communications}
As shown in Fig. \ref{usecases}, a UAV can serve as a flying UE, an aerial BS, or a flying relay.
\subsubsection{Flying UE}
By offering flexible and affordable options, UAVs have transformed many sectors. In filmmaking and sports, UAVs capture dynamic sights previously only possible with helicopters. Amazon and DHL are testing UAVs for small package deliveries to reduce traffic and personnel. In inaccessible locations where larger vehicles cannot operate, UAVs can acquire high-resolution data for 3D mapping and disaster response \cite{shahzadi2021uav}. Infrared sensors monitor crop health in agriculture, while thermal imaging helps find lost people and monitor criminal activity. UAVs also support wildlife conservation by deterring poachers, observing endangered species, and contributing to extreme weather predictions. Additionally, they enhance public safety by monitoring large gatherings, detecting illegal activities, and even providing entertainment through drone tournaments.

\subsubsection{Aerial BS}
Interactive multimedia applications have increased connectivity and coverage needs as technology progresses. Massive multiple input multiple outputs (mMIMO), heterogeneous networks, mmWave communication, and orthogonal multiple access systems have been developed to fulfill these demands, but each has limitations. In hard-to-reach places or high-density events like sports contests, UAV-assisted BS can boost future networks' capacity and coverage quickly and cheaply. They are perfect for on-demand services due to their speedy deployment and low cost. The portability and inexpensive cost of UAV-based aerial BS make it a lifeline for search and rescue missions in natural disasters, where communication infrastructure often fails. 

\subsubsection{Flying Relay}
UAVs can form a network among themselves through direct communication without the requirement for a central control system. This network operates like a flying ad-hoc network. UAV relay networks are highly efficient in ensuring dependable communication in inaccessible regions, whether because of physical barriers or during calamity. They greatly enhance the communication range, enabling connectivity to remote target regions that would otherwise be unattainable. The UAV relay nodes, which are deployed at a distance, gather and transmit data to a nearby BS. Due to the dependence of UAV communication on a direct line-of-sight (LoS) route, it is common to employ numerous UAVs to establish LoS communication when barriers are present. UAV relay networks have practical applications in various industries, such as border surveillance, monitoring disaster areas, remote sensing, and detecting forest fires.

\begin{table}[!t]
\renewcommand{\arraystretch}{1.2} % Optional: adjust cell height
\caption{Diagonal versus beyond diagonal RIS.}
\label{tab:comparison}
\centering
%\begin{adjustbox}{}
\begin{tabular}{|c|c|c|}
\hline
\textbf{Comparison} & \textbf{Diagonal RIS} & \textbf{ Beyond Diagonal RIS}  \\
\hline\hline
Signal Manipulation & Phase & Phase and Amplitude  \\
\hline
Adaptability & Low & High  \\
\hline
Beamforming capability & Low & High  \\
\hline
Energy Efficiency & High & Medium  \\
\hline
Circuit Complexity & Low & High  \\
\hline
Performance & Low & High  \\
\hline
Coverage & Limited & Enhanced  \\
\hline
Interference Mitigation & Low & High  \\
\hline
Spectral Efficiency & Low & High \\
\hline
Capacity Improvement & Low & High  \\
\hline
Designing Cost & Low & Medium \\
\hline
Power Requirement & Low & Medium  \\
\hline
\end{tabular}
%\end{adjustbox}
\end{table}
\section{Beyond Diagonal RIS Fundamentals}\label{SEC:III}
In this section, we first discuss the difference between D-RIS and BD-RIS, then we describe different hardware architectures of BD-RIS, and finally, we provide various configurations of UAV-based BD-RIS systems.
\subsection{D-RIS Versus BD-RIS}
The conventional D-RIS uses grid-arranged elements to independently manipulate the phase of incoming signal for passive beamforming, signal enhancement, and interference management. This model is simple and easy to deploy but limits control and flexibility \cite{10396846}. On the contrary, BD-RIS offers more advanced control mechanisms to adjust reflected signal amplitude, phase, and possibly polarization. This allows for advanced beamforming, multi-functional operations, and enhanced communication capabilities. Although the design and control of BD-RIS are more complicated and may demand high complexity compared to D-RIS, it excels in complex communication situations and offers higher performance and versatility compared to its counterpart, which is only suitable for simple communication scenarios \cite{10158988}. A detailed comparison is provided in Table \ref{tab:comparison}.

%%%%%%%%%%%%%%%%
\begin{figure}[!t]
\centering
\includegraphics[width=0.48\textwidth]{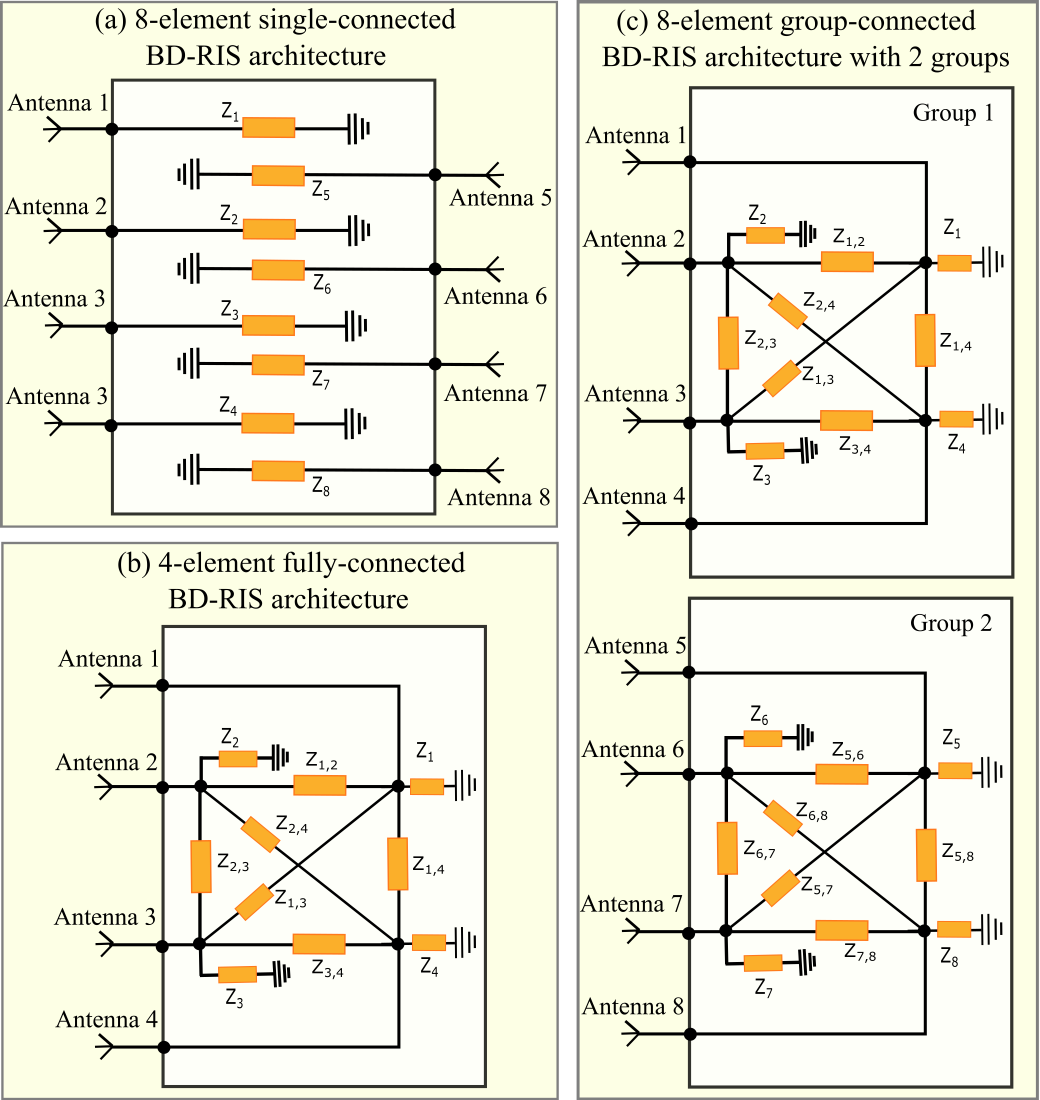}
\caption{BD-RIS hardware architectures include a 4-element single-connected, a 4-element fully-connected, and an 8-element group-connected configuration with a group size of 2. }
\label{Archi}
\end{figure}
%%%%%%%%%%%%%
\begin{table*}[!t]
\renewcommand{\arraystretch}{1.2} % Optional: adjust cell height
\caption{Characteristics of different BD-RIS architectures and modes.}
\label{architecture}
\centering
%\begin{adjustbox}{}
\begin{tabular}{|c|c|c|c|}
\hline
\textbf{Characteristics} & \textbf{Single-connected architecture} & \textbf{Fully-connected architecture} & \textbf{Group-connected architecture}  \\
\hline\hline
Number of groups & $K$ & 1 & $L$ \\
\hline
Group dimension & 1 & $K$ & $\bar{K}$ \\
\hline
Elements/group & 1 & $K^2$ & $\bar{K}^2$\\
\hline
Number of non-zero elements & $K$ & $K^2$ & $L\bar{K}^2$ \\
\hline
Reflective mode & $|\phi_{r,k}|^2=1$ & ${\bf \Phi}^H_{r}{\bf \Phi}_{r}={\bf I}_K$ & ${\bf \Phi}^H_{r,l}{\bf \Phi}_{r,l}={\bf I}_{\bar{K}}$  \\
\hline
Transmissive mode & $|\phi_{t,k}|^2=1$  & ${\bf \Phi}^H_{t}{\bf \Phi}_{t}={\bf I}_K$ & ${\bf \Phi}^H_{t,l}{\bf \Phi}_{t,l}={\bf I}_{\bar{K}}$  \\
\hline
Hybrid mode &$|\phi_{r,k}|^2+|\phi_{r,k}|^2=1$ & ${\bf \Phi}^H_{r}{\bf \Phi}_{r}+{\bf \Phi}^H_{t}{\bf \Phi}_{t}={\bf I}_K$  & ${\bf \Phi}^H_{r,l}{\bf \Phi}_{r,l}+{\bf \Phi}^H_{t,l}{\bf \Phi}_{t,l}={\bf I}_{\bar{K}}$ \\
\hline
\end{tabular}
%\end{adjustbox}
\end{table*}
%%%%%%%%%%%%%%%%
\begin{figure*}[!t]
\centering
\includegraphics[width=0.99\textwidth]{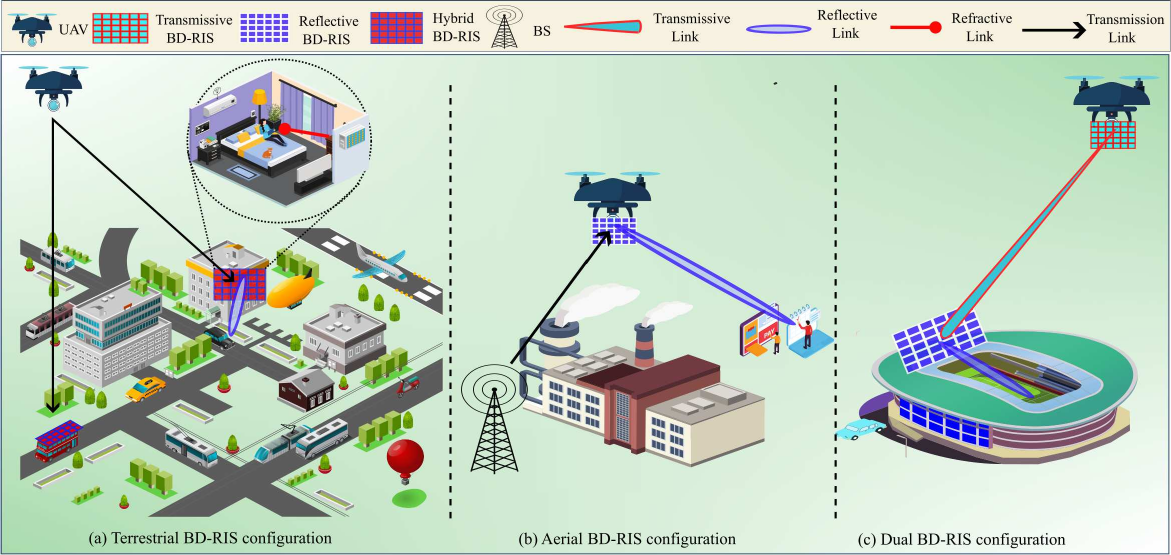}
\caption{Configuration of UAV-based BD-RIS systems: (a) Terrestrial BD-RIS configuration, (b) Aerial BD-RIS configuration, and (c) Dual BD-RIS configuration. }
\label{BDRISconfig}
\end{figure*}
%%%%%%%%%%%%%
\subsection{Hardware Architecture of BD-RIS}
Based on hardware architecture, BD-RIS can be categorized into single-connected, fully-connected, and group-connected types \cite{9913356}. Each architecture can operate in three modes: reflective, transmissive, and hybrid. This results in nine possible configurations, which are summarized in Table \ref{architecture}. In the following, we discuss each hardware architecture and its phase shift matrices when operating in the different modes in detail.
\subsubsection{Single-connected BD-RIS}
In this BD-RIS architecture, the reconfigurable elements on the surface are not interconnected and independently control the direction of the incident signal.
The phase response matrices in this architecture are all diagonal such as ${\bf \Phi}=\text{diag}(\phi_{1},\phi_{2}\dots\phi_{K})$ with $K$ is the total phase response elements. More specifically, the entries of single-connected BD-RIS should satisfy $|\phi_{r,k}|^2=1$ when operating in reflective mode, $|\phi_{t,k}|^2=1$ when operating in transmissive mode, and $|\phi_{r,k}|^2+|\phi_{r,k}|^2=1$ when operating in hybrid mode. Note that the hybrid mode is also called simultaneously transmitting and reflecting or intelligent omni-surfaces. Fig. \ref{Archi}(a) illustrates a 8-element single-connected hybrid BD-RIS architecture, where all 8 elements are independent and not inter-connected.
\subsubsection{Fully-connected BD-RIS}
In this BD-RIS architecture, the phase response elements on the surface are interconnected through reconfigurable impedance components. The elements do not operate independently but collectively control the direction of the incident signal. In fully-connected BD-RIS, the phase response matrices are all full matrices. More specifically, the matrices of fully-connected architecture should satisfy ${\bf \Phi}^H_{r}{\bf \Phi}_{r}={\bf I}_K$ when operating in reflective mode, ${\bf \Phi}^H_{t}{\bf \Phi}_{t}={\bf I}_K$ when operating in transmissive mode, and ${\bf \Phi}^H_{r}{\bf \Phi}_{r}+{\bf \Phi}^H_{t}{\bf \Phi}_{t}={\bf I}_K$ when operating in hybrid mode. A 4-element fully-connected reflective BD-RIS architecture is described in Fig. \ref{Archi}(b), where all 4 elements are interconnected, resulting in a full matrix.
\subsubsection{Group-connected BD-RIS}
In group-connected BD-RIS, the reconfigurable elements on the surface are divided into multiple groups to reduce system topology complexity. Note that in group-connected BD-RIS, each group is fully connected while ${\bf \Phi}_r$ and ${\bf \Phi}_t$ are all block diagonal, where ${\bf \Phi}_r=\text{blkdiag}({\bf \Phi}_{r,1},{\bf \Phi}_{r,2},\dots,{\bf \Phi}_{r,L} )$ and ${\bf \Phi}_t=\text{blkdiag}({\bf \Phi}_{t,1},{\bf \Phi}_{t,2},\dots,{\bf \Phi}_{t,L} )$ with $L$ is the number of groups. Suppose the number of elements in each group is the same such as $\Bar{K}=K/L$. In that case, the matrices of each group should satisfy ${\bf \Phi}^H_{r,l}{\bf \Phi}_{r,l}={\bf I}_{\bar{K}}$ when operating in reflective mode, ${\bf \Phi}^H_{t,l}{\bf \Phi}_{t,l}={\bf I}_{\bar{K}}$ when operating in transmissive mode, and ${\bf \Phi}^H_{r,l}{\bf \Phi}_{r,l}+{\bf \Phi}^H_{t,l}{\bf \Phi}_{t,l}={\bf I}_{\bar{K}}$ when operating in hybrid mode. Fig. \ref{Archi}(c) shows an 8-element group-connected reflective BD-RIS architecture, where the number of groups is 2.

\subsubsection{Architectural Trade-off}
The best-performing BD-RIS architecture is the fully-connected architecture, as it provides advanced beamforming capabilities, supports multiple users simultaneously, and achieves high spectral and energy efficiency. However, it has the highest complexity due to intricate interconnections and control requirements. On the other hand, the single-connected architecture is the least complex but offers limited performance. The group-connected architecture balances performance and complexity, making it a practical choice for many applications. In addition, BD-RIS can function as a transmitter by actively modulating signals, as a reflector/refractor to enhance coverage and mitigate blockages, or, theoretically, as a receiver by processing and absorbing incoming signals. While BD-RIS is predominantly used as a transmitter or reflector/refractor in communications, its application as a receiver is less explored and might be relevant in advanced system designs like BD-RIS-assisted MIMO networks.

\subsection{Different Configurations of UAV-Based BD-RIS Systems}
UAV-based BD-RIS systems can be configured in various ways to optimize communication performance. These configurations can be broadly classified into terrestrial, aerial, and dual setups, each offering unique advantages.

\subsubsection{Terrestrial BD-RIS Configurations}

Terrestrial BD-RIS units are installed on buildings or other fixed structures to reflect/refract signals between UAVs and ground stations for air-to-ground (A2G) and ground-to-air (G2A) links. This setup is particularly effective in urban areas with propagation obstacles, where strategically placed terrestrial BD-RIS units can significantly enhance coverage and signal strength by leveraging existing infrastructure. 

\subsubsection{Aerial BD-RIS Configurations}

Aerial BD-RIS units are mounted directly on UAVs, providing mobile signal reflection and beamforming capabilities. This configuration allows UAVs to dynamically adjust their positions and orientations to maintain optimal communication links, offering flexibility and adaptability in varying operational environments. Aerial BD-RIS configurations also utilize adaptive phase shifts, where the phase angles of aerial BD-RIS elements are continuously adjusted based on real-time communication needs and environmental conditions. This adaptive strategy ensures that signal reflection is consistently optimized, enhancing communication quality and efficiency across different scenarios.

\subsubsection{Dual BD-RIS Configurations}
Dual or hybrid configuration integrates both terrestrial-based and aerial BD-RIS units to achieve comprehensive coverage and flexibility. This setup adapts to diverse operational requirements by combining the stability of ground-based units with the mobility of UAV-mounted units, ensuring robust communication support in various environments. In collaborative configurations, multiple BD-RIS units work together to enhance signal paths and distribute computational loads efficiently. By leveraging the strengths of both terrestrial and non-terrestrial BD-RIS units, these configurations create a network of reflective surfaces that optimize signal propagation and computational performance, resulting in superior overall communication efficiency.
    
%\subsection{Challenges}

\section{Potential of BD-RIS and UAVs Integration}\label{SEC:IV}
This section explores the key advantages of integrating BD-RIS and UAV technologies in 6G NTNs.
\subsection{Transmissive BD-RIS mounted Transceiver}
Transmissive BD-RIS technology extends beyond conventional wireless communication by offering substantial potential for both transmitter and receiver applications, thereby paving the way for two distinct research avenues. In contrast to traditional antenna array UAV systems, BD-RIS-mounted UAV transceivers require fewer RF chains, leading to a simplified system architecture that significantly reduces both power consumption and operational costs. Additionally, transmissive BD-RIS overcomes the limitations seen in reflective BD-RIS transmitter architectures, such as feed blockage and self-interference \cite{10308579}. This approach enhances bandwidth and aperture efficiency, making it a more effective solution for advanced communication systems.

\begin{figure*}[!t]
\centering
\begin{subfigure}{.5\textwidth}
  \centering
  \includegraphics[width=1\linewidth]{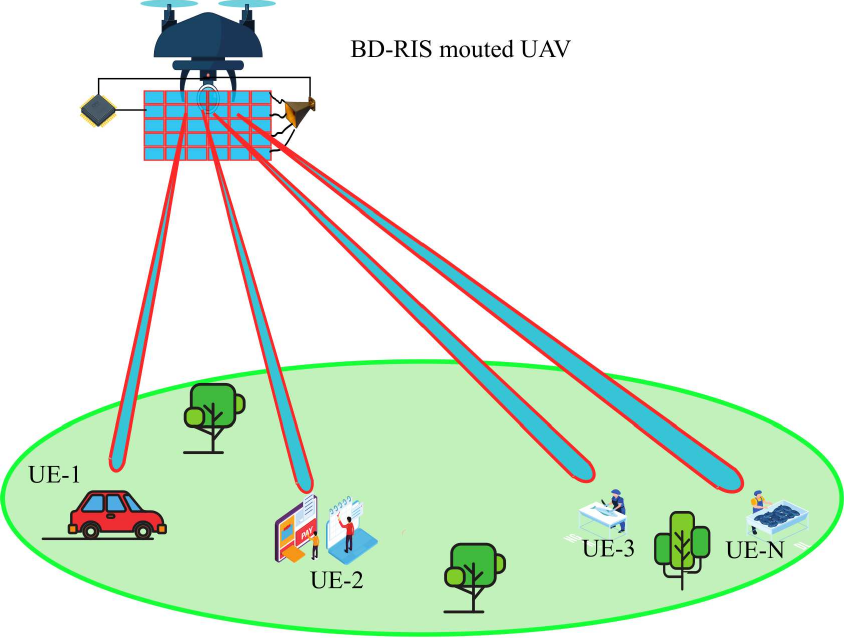}
  \caption{}
  \label{fig:sub1}
\end{subfigure}%
%\begin{subfigure}{.33\textwidth}
%  \centering
%  \includegraphics[width=1\linewidth]{Result1}
%  \caption{}
%  \label{fig:sub2}
%\end{subfigure}%
\begin{subfigure}{.5\textwidth}
  \centering
  \includegraphics[width=1\linewidth]{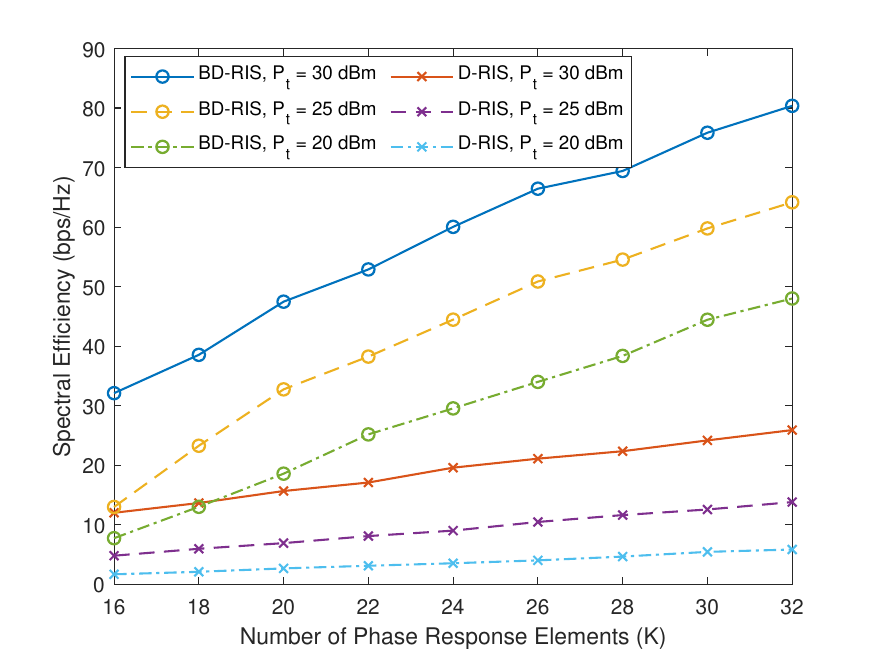}
  \caption{}
  \label{fig:sub3}
\end{subfigure}
\caption{(a) System model of transmissive BD-RIS mounted UAV communications, %(b) Spectral efficiency versus varying transmit power, 
(b) Achievable spectral efficiency of the system versus varying phase response elements.}
\label{fig:test}
\end{figure*}
\subsection{Extended Battery Life and Energy Efficiency}
The limited battery life of UAVs influences their operational lifetime and energy availability for communication. Extending operational time while carrying out necessary activities depends on effective energy management. Energy-efficient technologies guarantee UAVs may continue operations longer and successfully perform their tasks. Combining BD-RIS with UAVs will increase battery life and help to better manage energy. By use of signal reflection and transmission and enhancement of reception, BD-RIS can improve signal coverage and, therefore, lower the demand for high transmission power from UAVs. This energy-sustainable method enables UAV operations to last longer and aid in lower power consumption.
\subsection{Improved Interference Management}
In cases where UAVs run in the same frequency ranges as other equipment, co-channel interference might compromise communication quality. Maintaining dependable and clear channels of communication depends on controlling and reducing such interference. Furthermore, in settings with several UAVs, interference between their signals can become a significant problem. In situations where UAVs run in the same frequency ranges as other devices, BD-RIS can reduce co-channel interference. Advanced beamforming enables BD-RIS to assist in guiding interference away from UAV communication routes. BD-RIS can provide customized signal routes that minimize interference between multiple UAVs in environments, therefore guaranteeing accurate and dependable communication channels. 
\subsection{Strengthened Security and Privacy}
Security is a big issue since UAV communications are prone to cybersecurity problems like possible hacking and data leaks. The protection of private data and guarantee of operational integrity depend on strong encryption and safe communication methods being used. Privacy problems also surface primarily in relation to surveillance or monitoring UAVs are employed for. BD-RIS can improve the security and privacy of UAVs by helping to maximize signal pathways, which helps to lower the possibility of eavesdropping or interception. Furthermore, the BD-RIS can be made to incorporate security elements safeguarding data flow between UAVs and the network, therefore addressing possible cybersecurity risks and safeguarding privacy.

\section{Case Study: Transmissive BD-RIS Mounted UAV Communications}\label{SEC:V}
In this section, we present a technical case study that encompasses the system model, power consumption formulation, proposed solution methodology, and the corresponding numerical results.

\subsection{System Model and Proposed Solution}
We consider a downlink communication, where a $K$ element fully-connected transmissive BD-RIS mounted UAV transmits signals to $N$ UEs using $M$ resource blocks (RBs), as illustrated in Fig. \ref{fig:test}(a)\footnote{While it is true that the other two modes of BD-RIS (reflective and hybrid) are typically used to enhance communication when the line-of-sight (LoS) link is obstructed, transmissive BD-RIS can control and optimize the signal propagation actively, even when the LoS is available \cite{9133266}. Moreover, reflective and hybrid modes cause self-interference as the transmitter and receiver are located on the same side. The transmissive BD-RIS does not cause self-interference as the transmitter and receiver are on the opposite side of BD-RIS \cite{10242373}.}. It is important to note that using transmissive BD-RIS mounted on a UAV does not require complex signal processing as it is equipped with a UAV through a feed antenna. This makes transmissive BD-RIS an efficient technology for UAVs (due to limited energy resources on board) compared to a traditional multi-antenna transmitter, which consists of a large antenna array equipped with energy-hunger RF chains. 
%Let $\boldsymbol{\Phi}\in\mathbb C^{K\times K}$ be the phase shift matrix of transmissive BD-RIS such that $\boldsymbol{\Phi}\boldsymbol{\Phi}^H={\bf I}_K$, where ${\bf I}_K$ is the identity matrix. 
This framework assumes that the channel state information of UEs is available at UAV. The channel model between UAV and UEs follows Rician-distributed, which contains both LoS and non-LoS components.
%More specifically, the LOS component for each UE can be expressed as $[1,e^{-j\rho \sin{\theta}_i\cos{\varphi}_n},\dots, e^{-j\rho \sin{\theta}_n\cos{\varphi}_n(K-1)}]^T\otimes [1,e^{-j\rho \sin{\theta}_n\cos{\varphi}_n},\dots, e^{-j\rho \sin{\theta}_n\cos{\varphi}_n(K-1)}]^T$ with $\rho=\frac{2\pi f_c d_0}{c}$, where $c$ is the speed of light, $f_c$ is the carrier frequency, and $d_0$ denotes the BD-RIS element spacing. Furthermore, $\theta$ is the vertical and $\varphi$ is the horizontal angle of departure from UAV to UE-$n$. The non-LOS component can be derived as independent and identical such as non-LOS$\bf\sim\mathcal C(0,1)$. 

The proposed framework aims to maximize the spectral efficiency of transmissive BD-RIS mounted UAV communication by simultaneously optimizing the power allocation of UAV and phase shift defining of transmissive BD-RIS. This can be achieved by solving the joint optimization problem of spectral efficiency maximization. Due to the presence of the logarithmic function and the interdependence of the optimization variables, finding a direct joint solution is challenging. To overcome this, we employ an alternating optimization approach, iteratively obtaining an efficient solution using the gradient descent method. Specifically, the joint optimization problem is decoupled into two subproblems: 1) optimizing the power allocation at the UAV, given a fixed phase shift design at the transmissive BD-RIS, and 2) optimizing the phase shift design at the transmissive BD-RIS, given an obtained power allocation at the UAV. 

\subsection{Numerical Results and Discussion}
The numerical results of the proposed framework are obtained based on an average of 100 realizations. We also consider the conventional D-RIS as a benchmark framework to check the performance of the proposed transmissive BD-RS UAV communications. Unless mentioned otherwise, in the simulation, we consider the number UEs to be $N=10$, the number of RBs is $M=10$, the transmit power of UAV to be $P_t=20,25,30$ dBm, the noise variance is $\sigma^2=N_0\times B$ with $N_0=0.001$ mW/Hz, the carrier frequency is $f_c=28$ GHz, bandwidth over each RB is $B=20$ MHz, a UE can access one RB at given time, the number of phase response elements is $K=32$, and the Rician factor is $\eta_n=5$ dBm.

%In Fig. \ref{fig:sub2}(b), we plot the spectral efficiency against the varying transmission power of the UAV. We can observe that the transmit power of the UAV plays a crucial role in the system performance of the proposed BD-RIS and D-RIS mounted UAV communication. The spectral efficiency increases as the available transmit power of the UAV increases. We can also notice that the proposed BD-RIS framework achieves significantly high spectral efficiency compared to the conventional D-RIS framework. Accordingly,
Fig. \ref{fig:test}(b) investigates the spectral efficiency against the number of phase response elements. In this figure, we consider different power budgets of UAVs, i.e., 20 dBm, 25 dBm, and 30 dBm. It is interesting to notice that the phase response elements play a crucial role in the system performance of the proposed BD-RIS and D-RIS mounted UAV communication. It can also be seen that the performance of both frameworks is enhanced as the number of elements increases. However, the proposed BD-RIS mounted UAV communications achieve higher spectral efficiency than its counterpart D-RIS framework. For example, considering an equal system setup when the number of phase response elements is $K=24$, and the $P_t=30$ dBm, the proposed BD-RIS framework achieves 60 b/s/Hz while the benchmark D-RIS framework can only achieve 20 b/s/Hz.  This shows the effectiveness of the BD-RIS compared to the conventional D-RIS when operating in transmissive mode.   

\section{Future Research Directions}\label{SEC:VI}
The combination of BD-RIS with UAVs offers several exciting research opportunities that can greatly improve the performance and capabilities of future 6G networks. The following directions delineate prospective areas for future research.
\subsubsection{Advance Phase Shift Design and Optimization}
Future research can focus on the optimization of BD-RIS that goes beyond traditional linear phase shift design. The hybrid phase shift design of BD-RIS can provide enhanced control over electromagnetic waves, resulting in improved performance in complex propagation environments. Furthermore, it is worth investigating how artificial intelligence can be used to optimize dynamic BD-RIS. Machine learning algorithms can enable real-time adjustment of BD-RIS parameters, adapting to changing conditions to effectively optimize signal reflection and transmission.
\subsubsection{Cooperative Control of UAV-BD-RIS}
Another potential research direction is optimizing both the configurations of BD-RIS and the trajectories of UAVs together. Simultaneously optimizing the settings of the RIS and the trajectories of UAVs can significantly enhance communication performance. Implementing real-time adaption algorithms for UAV placements and RIS positions, based on instantaneous channel information and user mobility patterns, can significantly improve the responsiveness and reliability of the system.
\subsubsection{Physical Layer Security}
Enhancing physical layer security using BD-RIS can protect UAV communication links from intercepting and jamming attacks. Future research can investigate efficient methods for creating controlled multipath effects to confuse eavesdroppers and form secure channels that are difficult to intercept. Additionally, secure data transmission algorithms that leverage BD-RIS to create highly directional beams, reducing interception risks, should be designed.
\subsubsection{UAV-based Multi-BD-RIS networks}
Coordinating multiple BD-RIS systems deployed in a certain area to work collaboratively with UAVs can further enhance the overall system performance.  Moreover, the distributed architectures where BD-RIS systems and UAVs are strategically placed to maximize coverage and minimize interference are also promising. These studies should explore how different configurations affect system performance and scalability.
\subsubsection{Integration with Other Technologies}
Integrating UAVs and BD-RIS systems with 6G networks can improve capabilities, especially in delivering dependable coverage in inaccessible regions and supporting high-speed mobile customers. Research can investigate the utilization of UAV-BD-RIS systems in other technologies such as IoT networks to enhance connectivity for numerous low-power devices, joint communication, sensing and positioning, mobile edge computing, and THz communications. The merging of UAV-BD-RIS with these emerging should be studied.

\section{Conclusion}\label{SEC:VII}
In this work, we introduced the integration of BD-RIS and UAV communications within 6G NTNs. The study began by covering the essential aspects of UAV communications and the core principles behind BD-RIS technology. We then analyzed the potential benefits of combining BD-RIS with UAV systems in 6G NTNs, highlighting improvements in communication efficiency and system design simplification. A case study focused on transmissive BD-RIS mounted UAV communications, demonstrating its advantages over conventional D-RIS mounted UAV systems through numerical comparisons. The finding showed the benefits of BD-RIS mounted UAV communication compared to its counterpart D-RIS system. Lastly, we outlined key future research directions, including developing advanced phase shift design and optimization techniques, cooperative control of UAV-BD-RIS systems, enhancements in physical layer security, exploration of UAV-based multi-BD-RIS networks, and integration with other evolving technologies. These research areas present significant opportunities for optimizing performance, ensuring secure communication, and enhancing scalability in 6G systems.

%\bibliographystyle{IEEEtran}
\bibliographystyle{IEEEtran}% This is IEEEtran.bst file
\bibliography{Wali_EE}

\section*{Biographies}\small
\noindent {\bf Wali Ullah Khan [M]} (waliullah.khan@uni.lu) received a Ph.D.
degree in information and communication engineering from
Shandong University, China, in 2020. He is currently
working with the SIGCOM Research Group, SnT, University of Luxembourg. His research interests include 6G wireless communications, satellite communications, integrated terrestrial and non-terrestrial networks.
\vspace{0.2cm}

\noindent {\bf Eva Lagunas [SM]} (eva.lagunas@uni.lu) received a Ph.D. degree
in telecommunications engineering from the Polytechnic University of Catalonia (UPC), Barcelona, Spain, in 2014. She currently holds a research scientist position in the SIGCOM Research
Group, SnT, University of Luxembourg.
\vspace{0.2cm}

\noindent {\bf Asad Mahmood [S]} (asad.mahmood@uni.lu) received
his Master degrees in Electrical Engineering from COMSATS University Islamabad, Wah Campus, Pakistan, in 2020, and PhD degree in Computer Science from University of Luxembourg, Luxembourg in 2025. He is currently working as a Research Associate with the SIGCOM Research Group, SnT, University of Luxembourg.
\vspace{0.2cm}

\noindent {\bf Muhammad Asif} (masif@ujs.eu.cn) received a Ph.D. degree in information and communication engineering from the University of Science and Technology of China (USTC), Hefei, China, in 2019. He is currently working as a post-doctoral researcher at the School of Computer Science and Communication Engineering, Jiangsu University, Zhenjiang, China. His research interests include 6G wireless communications, and integrated terrestrial and non-terrestrial wireless networks.
\vspace{0.2cm}

\noindent {\bf Manzoor Ahmed} (manzoor.achakzai@gmail.com) received a
Ph.D. from Beijing University of Posts and Telecommunications China in 2015. He is currently a professor with the School of Computer and Information Science and also with the Institute for AI Industrial Technology Research, Hubei Engineering University, Xiaogan, China.
\vspace{0.2cm}

\noindent {\bf Symeon Chatzinotas [F]} (symeon.chatzinotas@uni.lu) received
Ph.D. degree in electronic engineering from the University of Surrey, Guildford, United Kingdom, in 2009. He is currently a full professor and head of the SIGCOM Research Group, SnT, University of Luxembourg.
%\newpage
%\include{ResponseFile}

\end{document}